\documentclass[pra,showpacs]{revtex4}
\usepackage{amssymb}
\usepackage{graphicx}
\usepackage{amsmath}
\usepackage{hyperref}
\usepackage{amsfonts}
\begin{document} 
\title{Non-locality breaking qubit channels}

\author{Rajarshi Pal }
\email{rajarshi@imsc.res.in}
\author{Sibasish Ghosh}
\email{sibasish@imsc.res.in}
\affiliation{Optics and Quantum Information Group, The Institute of Mathematical
Sciences, C. I. T. Campus, Taramani, Chennai 600113, India. }

\begin{abstract}
Entanglement breaking channels play a significant role in quantum information theory. In this work we investigate qubit channels through their property of `non-locality breaking', defined in a natural way but within the purview of  
CHSH nonlocality. This also provides a different perspective on the relationship between entanglement and nonlocality through the dual picture of quantum channels instead of through states. 
%This also provides a quantitative comparison between entagelment and non-loclaity through the dual picture of quantum channels instead of states.  
For a channel to be entanglement breaking  it is sufficient to `break' the entanglement of maximally entangled 
states. We provide examples to show that for CHSH nonlocality breaking such a property does not hold in general, though for certain channels and for a restricted class of states for all channels this holds.We also consider channels whose output remains local under SLOCC 
 and call them `strongly non-locality breaking'. 
 We provide a closed form necessary-sufficient condition for any two-qubit state to show    
hidden CHSH nonlocality, which is likely to be useful for other purposes as well. This in turn allows us to characterize all strongly non-locality breaking qubit channels.
It turns out that unital qubit channels breaking nonlocality of maximally entangled states are strongly non-locality breaking  while extremal qubit channels  cannot be so unless they are entanglement breaking.
\end{abstract}
\pacs{03.65.Bz,89.70.+c, 03.67.Mn, 42.50.Dv }

\maketitle
\section{Introduction:}

In quantum theory, physical processes correspond to trace-preserving completely positive(CPTP) maps and are also called quantum channels. 
Entanglement in state of a composite system is the primary resource for quantum information processing tasks and hence it is of considerable interest to study its behavior under local noise.
At an extreme end lies entanglement breaking channels  
which on acting on one side of any bipartite state makes the state separable. Entanglement breaking channels were studied for the first time in details in \cite{SH03} and have since played an important role  
in the theory of quantum information processing , for example in the theory of  capacities of quantum channels and so on \cite{SH02}.  
%As channels are isomorphically related to states via their action on a maximally entagled state(Choi-state) a natural question is if it is enough to fix attention on separability of the Choi-state of the channel. 

 In recent times it has been realized that nonlocality is also an information-theoretic resource and can be exploited for example to reduce the amount of communication needed in certain distributed 
computation tasks \cite{Bell-nonlocality-review}. In this work inspired by the notion of entanglement breaking(\cite{MBR03},\cite{SH03}) we look for a characterization of qubit channels based on their property of `non-locality breaking'. 
By seeing how  and to what extent  `non-locality breaking' is different from entanglement breaking we get additional insight into the relationship between entanglement and nonlocality. 
A state of a composite quantum system is known to be local iff the measurement statistics, 
for performing local measurements on the state, can be  simulated by a local hidden variable model. A `non-locality breaking channel' is thus one, which when acts on one subsystem of a composite system's state, 
brings every state to a local one.As {\emph{local}} states can be entangled \cite{Werner-89}, a basic question of interest is: 
{\textit{Can we have a non-entanglement breaking channel which when acting on one side of any bipartite state produces a state which has a local model?}}.
Before looking for an answer to the above question let us  note why this question is not already answered by the examples of mixed entangled states with local model (\cite{Werner-89}) that are  known. One of the main results from \cite{SH03}  
is that for a channel to be entanglement breaking it is necessary and sufficient for its dual state(in the Choi-Jamiolkowski sense) to be separable.It is apriori quite unclear if such a property holds good  for a `non-locality breaking channel'. 
Thus asking a channel to be `non-locality breaking' is a  stronger restriction than merely asking its dual-state to be local.

Further  entanglement cannot be increased by LOCC and  composition of an SLOCC map with an entanglement breaking channel  is again entanglement breaking. In the light of examples of genuine hidden quantum non-locality( \cite{Gen-hid-quan-nlc} ) this is also unlikely to hold for `non-locality breaking channels'           
and hence we  look at a stronger notion of non-locality breaking where not only are the output states of the channel required to be local, but they also do not show any hidden nonlocality under SLOCC. 
Such channels are said to be `strongly non-locality breaking'.
%In fact as  the output of the one-sided action of a channel on any bipartite pure state can be obtained by appplying a local filter to the Choi-state of the channel on the side whose reduction is maximally mixed, 
%hence for a channel to be non-locality breaking we not only need its Choi-state to have a local model but also all states which can be obtained from the Choi-state through action of a local filter on one 
%side(the side whose reduction is maximally mixed) should also have a local model.

%These are the directions we wish to pursue in this paper.  

It is known that a state of a composite system is local iff it satisfies all possible local realistic inequalities  \cite{PIT89} --- 
an impossibility to verify, in general.Also, the Bell-CHSH inequality(\cite{JSB65},\cite{CHSH}) is the only inequality for which the necessary-sufficient conditions for satisfaction by two-qubit states are known\cite{Hor95}. 
In this situation and as  a first study we  focus on qubit channels and the Bell-CHSH inequality.A qubit channel is henceforth said to be non-locality breaking if on acting on the qubit side of any qudit-qubit 
state it produces a state which satisfies a CHSH type inequality. Additionally, it is said to be `strongly non-locality breaking' if  the outputs satisfy the  same even after any SLOCC.   
We show that for both the notions it is enough to focus our attention on two-qubit pure state input. Also, we show that for non-locality breaking in the stronger sense it is enough to focus on local filtering through single qubit  filters and maximally  
entangled input.

We provide examples to show that for non-unital channels to be non-locality breaking it is $\emph{not}$ enough for the output state  of the channel for maximally entangled input to satisfy the Bell-CHSH inequality. 
In fact it may even not suffice to break the non-locality of all pure entangled states with a given Schmidt basis.  There
also seems to be exceptions like the amplitude damping channel which seems to be universally non-locality breaking for a certain parameter range. 
It is however true that channels breaking non-locality of a maximally entangled state also break that of all states whose reduction on the free side is maximally mixed.

Extending the work of Verstraete et al. in \cite{Ver} and \cite{Ver-lor} we provide a closed form necessary-sufficient condition for any two-qubit state to show hidden nonlocality \cite{Bell-nonlocality-review} with respect to the Bell-CHSH inequality.This is 
likely to be useful for other purposes as well.Using this we provide an exact characterization of all strongly non-locality breaking qubit channels. It turns out that  
unital qubit channels breaking nonlocality of a maximally entangled state are strongly non-locality breaking while  extremal qubit channels cannot be  
so unless  they are entanglement breaking.We exploit a recent example of genuine hidden nonlocality \cite{Gen-hid-quan-nlc} to show that a channel which {\emph{genuinely}} breaks the non-locality of maximally entangled states
(in the sense that its dual state has a local model) may not be strongly non-locality breaking. We also show numerically that the relative volume of entanglement breaking, non-locality breaking for maximally entangled states and strongly    
non-locality breaking channels in the six-dimensional real parameter space of all qubit channels is respectively about $0.24$, $0.81$ and $0.39$ . 

The paper is organized as follows. In section II  we introduce preliminary notions of    non-locality  and qubit channels. In section III we introduce notions of 'non-locality breaking' and show that for a channel to be  non-locality breaking or strongly non-locality 
breaking  it is sufficient to restrict to two-qubit pure state inputs.Additionally, we show that for non-locality breaking in the stronger sense it is enough to focus on local filtering operations through single qubit filters and maximally entangled input. 
In section IV we investigate conditions on qubit channels for breaking CHSH nonlocality of maximally entangled states. We provide
counterexamples to show that it is generally not enough to break non-locality of maximally entangled states for breaking non-locality of any input state. We  show numerically that channels of the amplitude damping form are an exception.  
We also show that channels breaking non-locality of maximally entangled states also break that of those states whose free sided reduction is maximally mixed. In section V we consider strongly non-locality breaking channels. We
first provide a closed-form necessary-sufficient condition for any two-qubit state to show hidden CHSH non-locality in a single copy scenario. Using this condition we then show  that amplitude damping channels or extremal qubit channels more  
generally cannot be strongly non-locality breaking unless they are entanglement breaking. We show that unital qubit channels breaking non-locality of maximally entangled states are strongly non-locality breaking .We  exploit a recent example of genuine hidden nonlocality in  literature to show the existence of a one parameter 
family of channels whose dual states (in the Choi-Jamiolkowski sense)  have local models but which are not strongly non-locality breaking. Moreover , we  numerically compute the relative volumes of  
entanglement breaking, non-locality breaking for maximally entangled states and strongly  non-locality breaking channels in the six-dimensional real parameter space of all qubit channels. We end in section VI with discussions.

\section{Preliminaries}
A qubit channel  $\Lambda$ is a trace preserving completely positive map (TPCP),
mapping $\mathcal{B}(\mathcal{C}^{2})$ the set of all bounded linear operators to itself.

In \cite{MB02} it was shown that any such map $\Lambda$ can
be written as, \begin{equation}
\Lambda(\rho)=U_{1}\circ\Lambda'\circ U_{2}(\rho)\label{map-composition}\end{equation}
 with $\Lambda'$ being a canonical TPCP map and $U_{1}$ and $U_{2}$
being unitary maps. If $\rho=\frac{1}{2}(I+x\sigma_{1}+y\sigma_{2}+z\sigma_{3})$
and $\rho'=\Lambda'(\rho)=\frac{1}{2}(I+x'\sigma_{1}+y'\sigma_{2}+z'\sigma_{3})$
then in the Bloch sphere representation the map $\Lambda'$ is given
by, \begin{equation}
\left[\begin{array}{c}
1\\
x'\\
y'\\
z'\end{array}\right]=\begin{bmatrix}1 & 0 & 0 & 0\\
t_{1} & \lambda_{1} & 0 & 0\\
t_{2} & 0 & \lambda_{2} & 0\\
t_{3} & 0 & 0 & \lambda_{3}\end{bmatrix}\left[\begin{array}{c}
1\\
x\\
y\\
z\end{array}\right],\label{map-canonical}\end{equation}
 with $t_{i}$ and $\lambda_{i}$ being real for all $i$ and satisfying conditions (given in \cite{MB02}) for complete positivity. Let , ${\mathbf{T}}_{\Lambda'}$ denote the $4 \times 4$ matrix in the above eqn. A channel $\Lambda$ is said  
to be unital if $\Lambda(I)=I$. For unital $\Lambda$ we have $\vec{t}=0$ in eqn. (\ref{map-canonical}).

\subsection{Nonlocality and the  Bell-CHSH inequality}
Suppose, for a composite system in the state $\rho_{AB}$ (acting on Hilbert space $H_A \otimes H_B$) shared between Alice and Bob, each party can perform measurements characterised by POVMs  $\sum_a M_{a|x} =I_A $ and $\sum_b M_{b|y} =I_B $ respectively
with indices $x$ and $y$ characterising the possible choices of measurement settings for each party.   
Then  $\rho_{AB}$  is said to be {\textit{local}} iff  the probability of obtaining outcomes $a$ and $b$ for measurement choices $x$ and $y$ for Alice and Bob  respectively can be written as, 
%$x,y$ index the set of measurements for $A$ and $B$), 
\begin{equation}
\label{locality-criterion}
Tr(M_{a|x} \otimes M_{b|y} \rho) = \int_{\Lambda} p_A(a|x,\lambda)p_B(b|y,\lambda)p(\lambda)d\lambda .
\end{equation}
Eqn. (\ref{locality-criterion}) reflects the fact that for `local' states the correlation between Alice's and Bob's outcomes for a certain choice of measurement settings can be completely 
{\textit{explained away}} by the `hidden variable' $\lambda$ (see \cite{Bell-nonlocality-review}) so that we have,
$p(ab|x,y,\lambda)=p(a|x,\lambda)p(b|y,\lambda)$.
In a typical Bell-inequality scenario one considers measurement settings $M_{a|x}$ for Alice and $M_{b|y}$ for Bob with $x,y \in \{1,2,..m\}$ and outcomes $a,b=1,2,...\Delta$. With the  choice of outcome labels 
$a,b \in \{-1,1\}$, $\Delta=2$ 
and expectation $\langle a_xb_y \rangle= \sum_{a,b} ab p(ab|xy))$ we have for a local state satisfying the condition in eqn.(\ref{locality-criterion}) the Bell-CHSH inequality given by,
\begin{equation}
 \langle a_0b_0 \rangle +  \langle a_0b_1 \rangle  +  \langle a_1b_0 \rangle  - \langle a_1b_1 \rangle  \leq 2 . \label{chsh-inequality}
\end{equation}

\subsection{Violation of Bell-CHSH inequality by  two-qubit states}

Consider the Hilbert space  $H=\mathcal{C}^2 \otimes \mathcal{C}^2 $ of a two-qubit system. Any state on $H$ can be represented using the Hilbert-Schmidt basis as follows:
\begin{equation}
\rho= \frac{1}{4}(I\otimes I  + \vec{r}.\vec{\sigma} \otimes I + I \otimes \vec{s}.\vec{\sigma} + \sum_{n,m=1}^3 t_{nm}\sigma_n \otimes \sigma_m )  ,
\end{equation}
with the coefficients $\vec{s},\vec{r} \in \mathcal{R}^3$ and  

\begin{equation}
\label{tmat}
t_{nm}=tr(\rho \sigma_n \otimes \sigma_m) 
\end{equation}

form a real $3 \times 3$ matrix which we shall denote by $T_{\rho}$. Here $I$ is the $2\times2$ identity matrix and $\sigma_1$, $\sigma_2$, $\sigma_3$ are Pauli spin matrices.  

Now, the Bell operator associated with the Bell-CHSH inequality (\ref{chsh-inequality})  has the following general form \cite{CHSH} :
\begin{equation}
\label{Bell-op}
B_{CHSH} = \hat{a}.\vec{\sigma} \otimes (\hat{b} + \hat{b'}).\vec{\sigma} + \hat{a'}.\vec{\sigma}  \otimes (\hat{b}-\hat{b'}).\vec{\sigma}, 
\end{equation}

where $\hat{a}$, $\hat{a'}$, $\hat{b}$, $\hat{b'}$ are unit vectors in $\mathcal{R}^3$. Then the Bell-CHSH inequality for $\rho $, following from the consideration  of local hidden variable theory,  is given by

\begin{equation}
\label{bell-ineq}
Tr({\rho}B_{CHSH}) \leq 2 .
\end{equation}

The matrix $U_{\rho} := T_{\rho}^T T_{\rho} $ is a symmetric one, and so it can be diagonalized. We denote the two greater (obviously non-negative) eigenvalues 
of $U_{\rho}$ by $u$ and $\tilde{u}$ . Then we define the quantity 
\begin{equation}
\label{M-value}
M(\rho) = u + \tilde{u}   .
\end{equation}

\textbf{Theorem.} Any two-qubit  density matrix $\rho$ violates inequality (\ref{bell-ineq}) for some operator of the form  (\ref{Bell-op}) (i.e.,for some choice of $\hat{a}$, $\hat{a'}$, $\hat{b}$ and  $\hat{b'})$ iff $M(\rho) > 1 $ .

 The optimal Bell violation for $\rho$ is given by $2\sqrt{M(\rho)}$.  For a proof of this theorem, see ref. \cite{Hor95}.  

\subsection{Hidden Bell-CHSH nonlocality}
\label{hidden-nonlocality-subsection}
Consider  a local filtering transformation taking any two-qubit state $\rho$ to another two-qubit state
\begin{equation}
\rho'=\frac{(A \otimes B)\rho (A^{\dagger} \otimes B^{\dagger})}{Tr(A^{\dagger}A 
\otimes B^{\dagger}B\rho) } \label{local-filtering-eq1}
\end{equation} . 
Then, $\rho$ is said to show hidden CHSH non-locality  iff $\rho'$ violates the Bell-CHSH inequality \cite{CHSH} for at least one choice of $A$,$B$.

 \section{Non-locality breaking and strongly non-locality breaking  channels}

A Bell inequality in a scenario consisting of $\Delta$ measurements per site  and $m$ outcomes per measurement  can be written as a 
hyperplane separation condition in a $t=2(\Delta-1)m + (\Delta-1)^2m^2$ dimensional subspace of ${\mathcal{R}}^{\Delta^2 m^2}$(the space of $\Delta^2m^2$ dimensional probability vectors $\vec{p}$ with components $p(ab|xy)$ )as,

\begin{equation}
\label{bell-in-gen-cond}
\underbar{s}. \underbar{p} = \sum_{abxy} s^{ab}_{xy} p(ab|xy) \leq S_k 
\end{equation}
with the inequality being satisfied by all probability vectors $\vec{p}$   satisfying eqn. (\ref{locality-criterion}) and violated otherwise(see \cite{Bell-nonlocality-review} for more details).

{\textbf{Defn.:}}

(i) A channel $\$_A:{\mathcal{B}}({\mathcal{C}}^d) \rightarrow {\mathcal{B}}({\mathcal{C}}^{d'}) $,   is said to be {\textit{non-locality breaking}}, if acting on  side A of \textit{any} bipartite  state $\rho_{BA}$, it produces a 
state $\rho'_{BA}= (I \otimes \$_A) (\rho_{BA})$ which satisfies an inequality/inequalities of the form of eqn. \ref{bell-in-gen-cond} .  

\vspace{5mm}
(ii)Again, it is said to be {\textit{strongly non-locality breaking}} if under any SLOCC operation $\Omega$ , $\Omega(\rho'_{BA})$ satisfies the same inequality/inequalities for an arbitrary choice of $\rho_{BA}$.

\vspace{10mm}

{\textbf{Remark:}} Both the definitions are w.r.t a set of inequalities like eqn. (\ref{bell-in-gen-cond} ) and for {\textit{genuine}}, non-locality breaking or non-locality breaking in the stronger sense 
one should consider the locality criterion( \ref{locality-criterion}).

\subsection{Qubit channels and CHSH nonlocality}
In this subsection we focus on qubit channels and non-locality breaking with respect to the Bell-CHSH inequality. 

We begin  by showing that for both the notions of non-locality breaking it is sufficient to consider only two-qubit pure states as input and single qubit filters. 
This follows from Lemma 1 and 2 proved below. 

Suppose we have a qubit channel $\$$ that breaks non-locality of arbitrary two-qubit states $|\alpha\rangle \in {\mathcal{C}}^2 \otimes {\mathcal{C}}^2 $ i.e., 
\begin{equation}
\label{pabxy}
p(ab|xy) = Tr((I \otimes \$)(|\alpha\rangle \langle \alpha|)(M_{a|x} \otimes M_{b|y})) ,
\end{equation} 
satisfies eqn. (\ref{bell-in-gen-cond}) for all $|\alpha\rangle$ and any qubit POVMs $M_{a|x}$ and $M_{b|y}$ for all $x,y \in \{1,2,...\Delta\}$ and $a,b \in \{1,2,..,m\}$.

The following Lemma tells us that the channel also breaks the non-locality of {\textit{any}} qudit-qubit state. 

\vspace{5mm}
{\textbf{Lemma 1:}} Let $\$$ be a qubit channel for which $p(ab|xy)$ given by eqn. (\ref{pabxy}) satisfies an inequality of the form (\ref{bell-in-gen-cond}) for an arbitrary two-qubit state $|\alpha\rangle$ and qubit POVMs 
$M_{a|x}$ and $M_{b|y}$ as described before. Then for any arbitrary  state 
$\rho \in  {\mathcal{B}}\left({\mathcal{C}}^d \otimes {\mathcal{C}}^2\right) $  and arbitrary set of  POVMs $\sum_a N_{a|x} = I_d$ and $\sum_b N_{b|y} = I_2$,    $p'(a,b|x,y) = \mbox{Tr}(((I \otimes \$)\rho) (N_{a|x} \otimes N_{b|y}))$ 
also satisfies  eqn. (\ref{bell-in-gen-cond}).
\vspace{5mm}

{\textbf{Proof}}:  

Consider any qudit-qubit state $|\beta\rangle  \in {\mathcal{C}}^d \otimes {\mathcal{C}}^2 $. Choosing the Schmidt basis for the state
so that $|\beta\rangle = \sqrt{\lambda}|e_0f_0\rangle + \sqrt{(1-\lambda)}|e_1f_1\rangle$ where $\{|e_0 \rangle,|e_1\rangle\}$ is a two-dimensional ONB in
${\mathcal{C}}^d $,  $\{|f_0 \rangle,|f_1 \rangle \}$ an ONB of the qubit  on which the channel acts and  $\lambda \in [0,1]$  we have 
$\rho_{\$,\beta}=(I \otimes \$)(|\beta\rangle \langle \beta|) \in {\mathcal{B}}\left({\mathcal{C}}^2 \otimes {\mathcal{C}}^2 \right)$ where
the first ${\mathcal{C}}^2$ is the space spanned by $\{|e_0 \rangle,|e_1\rangle\}$. Again, suppose $\sum_a N_{a|x} = I_d$ and $\sum_b N_{b|y} = I_2$ are two sets of POVMs
acting respectively on qudit and qubit systems.  The probability of clicking of $N_{a|x}$ and $N_{b|y}$ for the system in the state  
$\rho_{\$,\beta}$ is given by,
\begin{eqnarray}
p'(a,b|x,y) &=& \mbox{Tr}(\rho_{\$,\beta} (N_{a|x} \otimes N_{b|y})) \nonumber\\
&=& \mbox{Tr} ((P_2 \otimes I_2) \rho_{\$,\beta} (P_2 \otimes I_2) (N_{a|x} \otimes N_{b|y})), (\mbox{where  } P_2=|e_0\rangle \langle e_0| + |e_1\rangle \langle e_1|) \nonumber\\
&=& \mbox{Tr} ((P_2 \otimes I_2) \rho_{\$,\beta} (P_2 \otimes I_2) ((P_2 + (I_d-P_2))\otimes I)(N_{a|x} \otimes N_{b|y})) \nonumber \\
&=& \mbox{Tr} (\rho_{\$,\beta} ((P_2 N_{a|x}P_2) \otimes N_{b|y})) \nonumber \\
&=& \mbox{Tr} (\rho_{\$,\beta} (N'_{a|x} \otimes N_{b|y})) 
\end{eqnarray}
with $N'_{a|x} = P_2 N_{a|x}P_2 $ being a qubit POVM satisfying $\sum_a N'_{a|x} = I_2$. Thus $p'(ab|xy)$ with $x$ and $y$ indexing respectively a
set of qudit and qubit POVMs, obtained from  $|\beta\rangle$ through the action of $N_{a|x}$ and $N_{b|y}$ through eqn. (\ref{pabxy}) will 
satisfy the Bell inequality given by eqn. (\ref{bell-in-gen-cond}), due to $\$$ being non-locality breaking for all two-qubit pure
states. The arguments extend to  mixed qudit-qubit input states  through convexity.

\hfill $\square$

\vspace{3mm}

Let us move on to  non-locality breaking in the stronger sense. Here we only consider inequalities with two two-outcome measurement settings per site (i.e, eqn. (\ref{bell-in-gen-cond}) with $\Delta=m=2$ ).
 Let
\begin{equation}
\label{local-filter-output}
\rho' = \frac{(A \otimes B) ((I \otimes \$)(|\alpha\rangle \langle \alpha|)) (A^{\dagger} \otimes B^{\dagger})}{Tr((A^{\dagger}A \otimes  B^{\dagger}B) (I \otimes \$)(|\alpha\rangle \langle \alpha|))}
\end{equation}
with $\$$ and $|\alpha\rangle$ being a qubit channel and an arbitrary two-qubit
pure state respectively and $A,B $ being $2 \times 2$ complex matrices acting as   qubit  filters. Let  us suppose now that $p(ab|xy) = Tr(\rho' (M_{a|x} \otimes M_{b|y}))$ obtained through  qubit POVMs 
$\{M_{a|x}: a=1,2\}$, $\{M_{b|y}:b=1,2\}$  satisfies eqn. (\ref{bell-in-gen-cond}) for any $|\alpha\rangle$, $A$, $B$. The next Lemma shows that this condition ensures that $\$$ is strongly nonlocality breaking.

\vspace{5mm}
{\textbf{Lemma 2}:}
Let $\$$ be an arbitrary qubit channel for which $p(ab|xy)$ obtained from any $\rho'$ defined above satisfies  eqn. (\ref{bell-in-gen-cond})  with 
$\Delta=m=2$ i.e, two two-outcome measurement settings per site.  Then $\$$ is strongly non-locality breaking w.r.t such an inequality.   
\vspace{10mm}

{\textbf{Proof}}:
To begin with, let us show that locality of   $\rho'$ in eqn. (\ref{local-filter-output})  is strong enough to ensure locality of $\rho_1=(I \otimes \$)(|\alpha\rangle \langle \alpha|)$ for any SLOCC operation 
$\Omega: {\mathcal{B}}({\mathcal{C}}^2 \otimes {\mathcal{C}}^2 ) \rightarrow {\mathcal{B}}({\mathcal{C}}^2 \otimes {\mathcal{C}}^2 )$  .  Let, 
\begin{equation}
 \Omega(\rho_1)= \frac{\sum_k (A_k \otimes B_k)\rho_1 (A_k^{\dagger} \otimes B_k^{\dagger})}{Tr(\sum_k ( A_k^{\dagger} A_k \otimes B_k^{\dagger} B_k)\rho_1 )} \label{Krauss-rep} \end{equation} with $A_k,B_k $ being $2 \times 2$ complex matrices. Defining, 

$p_k= Tr(( A_k^{\dagger} A_k \otimes B_k^{\dagger} B_k)\rho_1 ))$ and $\rho'_k= (A_k \otimes B_k)\rho_1 (A_k^{\dagger} \otimes B_k^{\dagger})/p_k$ we have, $\Omega(\rho_1)= \frac{\sum_k p_k\rho'_k}{\sum_k p_k}$ . As this is a convex combination 
for $\Omega(\rho_1)$ to violate inequality (\ref{bell-in-gen-cond}) we must have
at least one $\rho'_k$ which violates inequality (\ref{bell-in-gen-cond}) for some choice of measurement setting. The aforesaid condition on  $\rho'$ in eqn. (\ref{local-filter-output}) guarantees that this does not happen for any $\rho'_k$. 

The general proof follows by contradiction . Let us assume that there is a SLOCC operation $\Omega': {\mathcal{B}}({\mathcal{C}}^2 \otimes {\mathcal{C}}^2 ) \rightarrow {\mathcal{B}}({\mathcal{C}}^d \otimes {\mathcal{C}}^d )$ 
under which $\rho_1=(I \otimes \$)(|\alpha\rangle \langle \alpha|)$ violates inequality (\ref{bell-in-gen-cond} with $\Delta=m=2$)
for some choice of measurement setting. Now by 
Result 2 of \cite{Mas-06} $\Omega'(\rho_1)$ can be transformed by a SLO ($\Omega_1$, say)to a two-qubit state which violates inequality  (\ref{bell-in-gen-cond} with  
$\Delta=m=2$ ) by an equal or larger amount. 
Thus $\Omega_1 \circ \Omega'=\Omega_2$ is an SLOCC for which 
$\Omega_2(\rho_1) \in {\mathcal{B}}({\mathcal{C}}^2 \otimes {\mathcal{C}}^2)$ violates inequality (\ref{bell-in-gen-cond}). As we saw in the previous paragraph, locality of $\rho'$ in eqn. (\ref{local-filter-output})  ensures that
this does not happen and hence we have a contradiction.  The argument can be generalized to consider qudit-qubit mixed 
input states using Schmidt decomposition and convexity as before. If there exists a deterministic LOCC for which our proposition is not true then by using the representation of that LOCC with separable superoperators, 
like in eqn . (\ref{Krauss-rep}), with the added restriction $\sum_k A_k^{\dagger}A_k \otimes B_k^{\dagger}B_k= I \otimes I $ 
we can construct a local filtering transformation with   one of the pairs $\left(A_k,B_k\right)$ for which our proposition 
will also be violated by convexity. 

\hfill $\square$ 

One of the most important properties of entanglement breaking channels proved  in \cite{SH03} is that for a channel acting on density operators of a $d$-dimensional system it is enough to look at 
separability of the dual state of the channel, $\rho_{\Phi^+,\$}= (I \otimes \$)(|\Phi^+\rangle \langle \Phi^+|) $,with $|\Phi^+\rangle =\frac{1}{\sqrt{d}} \sum_{i=0}^{d-1} | ii \rangle$. For non-locality breaking we have no such thing. But the following
Lemma shows that for ``strongly non-locality breaking channels'' within the purview of CHSH-nonlocality it is enough to check if the Choi-state of the channel
(i.e.,$\rho_{\Phi^+,\$}$) show any hidden non-locality under a single local filtering operation.

\vspace{5mm}
{\textbf{Lemma 3}:}  An arbitrary qubit channel is strongly non-locality breaking w.r.t an inequality of the form of eqn. (\ref{bell-in-gen-cond}) with $\Delta=m=2$ iff $\rho'$ defined in eqn. (\ref{local-filter-output})  satisfies 
(\ref{bell-in-gen-cond}) for arbitrary $A$, $B$ and 
$|\alpha\rangle =|\Phi^+\rangle= \frac{1}{\sqrt{2}} \sum_{i=0}^1|ii\rangle$.  
\vspace{5mm}

\textbf{Proof}:
The most general two-qubit pure state in the Schmidt form is given
by, $|\alpha\rangle=\sqrt{\lambda}|e_{1}f_{1}\rangle+\sqrt{1-\lambda}|e_{2}f_{2}\rangle=(U\otimes V)(\sqrt{\lambda}|00\rangle+\sqrt{(1-\lambda)}|11\rangle)$,
with $\lambda\in[0,1]$ and the $2\times2$ unitary matrices $U$
and $V$ being given by: $U|0\rangle=|e_{1}\rangle,V|0\rangle=|f_{1}\rangle,U|1\rangle=|e_{2}\rangle$
and $V|1\rangle=|f_{2}\rangle$ with $\{|0\rangle,|1\rangle\}$ being the standard ONB for ${\mathcal{C}}^2$.

For $\lambda\in[0,1]$, let \begin{equation}
W_{\lambda}=\sqrt{\lambda}|0\rangle\langle0|+\sqrt{(1-\lambda)}|1\rangle\langle1|.\label{Wlambda-filter}\end{equation}

Now, we have $(U \otimes V) (W_{\lambda} \otimes I ) |\Phi^+ \rangle = (\sqrt{\lambda} |e_1f_1 \rangle + \sqrt{1-\lambda} |e_2f_2 \rangle)/\sqrt{2} = \frac{1}{\sqrt{2}}|\alpha \rangle $.

Now using the facts that a qubit channel $\Lambda$ is a trace-preserving map and that for any operator $A$, $(I \otimes A)|\Phi^+\rangle = (A^T \otimes I)|\Phi^+\rangle  $ it is
easy to show that, \begin{eqnarray}
\rho_{\alpha,\Lambda} & = & (I\otimes\Lambda)(|\alpha\rangle\langle\alpha|)\nonumber \nonumber \\ 
 & = & \frac{(A_{1}\otimes I)\rho_{\Phi^{+},\Lambda}(A_{1}^{\dagger}\otimes I)}{Tr((A_{1}^{\dagger}A_{1}\otimes I)\rho_{\Phi^{+},\Lambda})},\label{choi-state-filter} \label{mes-enough}\end{eqnarray}
 with the filter $A_{1}=UW_{\lambda}V^{T}$ , $\rho_{\Phi^{+},\Lambda} = (I \otimes \Lambda)\left(|\Phi^+\rangle \langle \Phi^+|\right)$. 
 
As local filtering operations can be composed  it follows from Lemma 2 that in order 
to check if $\$$ is strongly non-locality breaking it is sufficient to check if $\rho_{\Phi^{+},\Lambda}$ show any hidden non-locality.  

\hfill $\square$ 

\vspace{5mm}
Henceforth, a  channel $\$_A$, acting on qubit A,  will be said to be  {\textit{non-locality breaking}}, if acting on  side A of \textit{any} two-qubit state $\rho_{BA}$, it produces a 
state $\rho'_{BA}= (I \otimes \$_A) (\rho_{BA})$ which satisfies the Bell-CHSH inequality, {\it{i.e.}}, 
we have

\begin{equation}
\label{nonloc-condition}
M(\rho') \leq 1 .
\end{equation}

%Further, a qubit channel $\$_A$ will be  said to be {\textit{strongly non-locality breaking}} if for {\textit{any}}  single qubit filters $A$, $B$,
%\begin{equation}
%\label{snlc-condition}
%M\left(\frac{(A \otimes B) \rho_{\Phi^{+},\$_A} (A^{\dagger} \otimes B^{\dagger})}{Tr( (A^{\dagger}A \otimes B^{\dagger} B) \rho')}  \right)  \leq 1 .
%\end{equation}

\section{Breaking nonlocality of maximally entangled states}
In this section we produce various examples to show that breaking non-locality of a maximally entangled state is typically not enough for a channel to be non-locality
 breaking for all states. 
 
 In the Hilbert-Schmidt basis , $\rho_{\Phi^{+},\Lambda'}$ with $\Lambda'$ being the canonical map in eqn. (\ref{map-canonical}) is given by,
\begin{equation}
\label{choi-statehs} 
\rho_{\Phi^+, \Lambda'}= \frac{1}{4}( I \otimes I + I \otimes \vec{t}.\vec{\sigma} + \lambda_1 \sigma_1 \otimes \sigma_1 - \lambda_2 \sigma_2 \otimes \sigma_2 +  \lambda_3 \sigma_3 \otimes \sigma_3) .
\end{equation}
Now from eqn. (\ref{choi-state-filter}) and (\ref{map-composition}) using  
$ (I \otimes A)|\Phi^+\rangle  = (A^T \otimes I)|\Phi^+ \rangle $   we have 
\begin{equation}
\rho_{\alpha,\Lambda} = 
(W_{\lambda}V^{T} \otimes I) \rho_{\Phi^+, \Lambda'} ((W_{\lambda}V^{T})^{\dagger} \otimes I) /tr((W_{\lambda}^{\dagger}W_{\lambda}\otimes I)
|\Phi^+\rangle \langle \Phi^+|)
\end{equation}
 where we have neglected stray local unitaries which do not affect the Bell-violation of $\rho_{\alpha,\Lambda}$ . We further have, on using $tr((W_{\lambda}^{\dagger}W_{\lambda}\otimes I)
|\Phi^+\rangle \langle \Phi^+|)=\frac{1}{2}$,

 \begin{equation}
\label{eqn1-rho1}
\rho_{\alpha,\Lambda}=  2(W_{\lambda} \otimes I)\frac{1}{4}( I \otimes I + I \otimes \vec{t}.\vec{\sigma} + \lambda_1 V^T \sigma_1 ({V^T})^{\dagger}\otimes \sigma_1 - \lambda_2  V^T \sigma_2  
({V^T})^{\dagger}  \otimes \sigma_2 +  \lambda_3 V^T \sigma_3  ({V^T})^{\dagger} \otimes \sigma_3) (W_{\lambda}^{\dagger} \otimes I).
\end{equation}

Now let $ V^T \sigma_i ({V^T})^{\dagger}= \sum_{j=1}^3 R_{ij}\sigma_j $ where $R= (R_{ij})$ is the real rotation matrix in three 
dimensions corresponding to the matrix $V^T \in SU(2)$. Now using the action of $W_{\lambda}$ on the basis elements $I, \sigma_i, i=1,2,3$, we 
have from eqn. (\ref{eqn1-rho1}) 

\begin{eqnarray}
\label{rho'}
\rho_{\alpha,\Lambda} =& \frac{1}{4} (I \otimes I + (2\lambda-1)\sigma_3 \otimes I + I \otimes \vec{t}.\vec{\sigma} + \lambda_1R_{13}(2\lambda-1)I \otimes \sigma_1 -  \lambda_2R_{23}(2\lambda-1)I \otimes \sigma_2 +  \nonumber \\
& \lambda_3R_{33}(2\lambda-1)I \otimes \sigma_3 + \sum_{ij=1}^3t'_{ij} (\sigma_i \otimes \sigma_j)) .
\end{eqnarray} .

The entries for the correlation matrix $t'$ for $\rho_{\alpha,\Lambda}$ are given by: 
\begin{equation}
\label{Trho'}
(T_{\rho_{\alpha,\Lambda}})_{ij} =  \begin{pmatrix}  2\sqrt{\lambda(1-\lambda)} \lambda_1R_{11} & -2\sqrt{\lambda(1-\lambda)}  \lambda_2 R_{21} & 2\sqrt{\lambda(1-\lambda)} \lambda_3R_{31} \\  
 2\sqrt{\lambda(1-\lambda)} \lambda_1R_{12} & -2\sqrt{\lambda(1-\lambda)}  \lambda_2 R_{22} & 2\sqrt{\lambda(1-\lambda)} \lambda_3R_{32} \\
 \lambda_1R_{13} + (2\lambda-1)t_1 & -\lambda_2 R_{23} + (2\lambda-1)t_2 &  \lambda_3R_{33} + (2\lambda-1)t_3  \\ \end{pmatrix}
\end{equation} 

$=\mbox{Diag}(\alpha',\alpha',1) \begin{pmatrix} R_{11} & R_{21} & R_{31} \\ R_{12} & R_{22} & R_{32} \\ R_{13} & R_{23} & R_{33} \end{pmatrix} \mbox{Diag}(\lambda_1,-\lambda_2, \lambda_3) + 
\begin{pmatrix} 0 & 0 & 0 \\ 0 & 0 & 0 \\ (2\lambda - 1)t_1 & (2\lambda - 1)t_2 &  (2\lambda - 1)t_3 \end{pmatrix} $

 with $\alpha'=2\sqrt{\lambda(1-\lambda)}$.
 
\subsection{Non-locality breaking condition for maximally entangled states} 
The condition for a channel to break non-locality of a maximally entangled state follows from eqn. (\ref{nonloc-condition}) by taking $R=I$ and $\lambda=\frac{1}{2}$ in eqn. (\ref{Trho'}) and is given by
\begin{equation}
\label{nlb-mes-1}
\lambda_1^2 +  \lambda_2^2 \leq 1, 
\end{equation}
assuming $\lambda_1 \geq  \lambda_2  \geq \lambda_3$.   
 
\subsection{Examples and counterexamples of universal non-locality breaking} 
In this subsection we provide three examples to show that breaking non-locality of maximally entangled state may or may not be sufficient to break non-locality of all states.

We choose the following channel parameters for  non-unital qubit channels of the canonical form (\ref{map-canonical}),

\vspace{5mm}
i)
\begin{equation} \label{parameters-1}  \lambda_1=\frac{1}{\sqrt{2}} , \lambda_2=\frac{1}{\sqrt{2}}, \lambda_3=\frac{1}{2}, 
t_1=-0.12,   t_2=0.047,   t_3=-0.210. \end{equation} 
 
 The channel with the above parameters saturates the non-locality breaking condition given by eqn.(\ref{nlb-mes-1}).
 
 We choose $\lambda=0.4$ (note $\lambda=0.5$ corresponds to the maximally entangled state). Now, for any $R \in SO(3)$ we can write it (using Euler angles) as \begin{equation} \label{euler} R=R_z(\alpha_0)R_y(\beta_0)R_z(\gamma_0), \end{equation} 
 with $\alpha_0,\beta_0,\gamma_0 \in [0,2\pi], 
R_z(\alpha_0)=\begin{pmatrix} cos(\alpha_0) & -sin(\alpha_0) & 0 \\ sin(\alpha_0) & cos(\alpha_0) &  0 \\  0 & 0 & 1 \end{pmatrix} \mbox{ and }  R_y(\beta_0)= \begin{pmatrix} cos(\beta_0) & 0 & sin(\beta_0) \\ 0 &  1 & 0 \\ 
-sin(\beta_0)&  0 & cos(\beta_0) \end{pmatrix}  $ .

We choose $\alpha_0=1.2 , \beta_0=1.4, \gamma_0=3.5$ ,{\it{i.e.}}, resp. about   $0.382 \pi$, $0.4456 \pi$ and $1.11 \pi$ . 

\vspace{10mm}

For this choice of $\lambda$ and $R$ and the channel parameters given in eqn. (\ref{parameters-1}) , we have for $\rho_{\alpha,\Lambda}$ in eqn. (\ref{rho'}),  $M(\rho_{\alpha,\Lambda}) =  1.01094$ .
Here $R$ is the $3\times3$ real rotation matrix corresponding to 
the $2\times2$ $SU(2)$ matrix $V^T$ appearing in the expression for $\rho_{\alpha,\Lambda}$ in eqn. (\ref{eqn1-rho1}). 
\vspace{5mm}

Thus clearly the channel does not  break the non-locality of the state arising from the action of $W_\lambda V^T \otimes I$ on the maximally entangled state $|\Phi^+ \rangle$ . However this channel does break the non-locality 
of all states of the form $\sqrt{\lambda}|00\rangle +  \sqrt{1-\lambda}|11\rangle $ as we have checked by taking $R=I$ and varying over  $\lambda$. Hence, breaking
non-locality of all entangled states with a given Schmidt basis is not enough.

ii)   \begin{equation} \label{parameters}  \lambda_1=0.7 , \lambda_2=0.71, \lambda_3=0.7, 
t_1=0.28,   t_2=0.01,   t_3=-0.1 \end{equation} .
 
We further choose $\lambda=0.45$ and $R=I$ and we have for $\rho_{\alpha,\Lambda}$ in eqn. (\ref{rho'}),  $M(\rho_{\alpha,\Lambda}) =  1.0159$. Thus a channel breaking
 non-locality of a maximally entangled state may not even break it for all states with a given Schmidt basis.

iii) The {\textit{amplitude damping channel}}: 
The  vectors $(\vec{t}$ and $\vec{\lambda})$ of the amplitude-damping channel $\Phi$ (as defined in eqn.(\ref{map-canonical})) are given 
respectively by $(0,0,p)$ and  ($\sqrt{(1-p)}, \sqrt{(1-p)}, (1-p)$) . Thus, to break non-locality of $|\Phi^+\rangle$ we must have from eqn. (\ref{nlb-mes-1}), 
$p \geq \frac{1}{2}$. 

The correlation matrix for the state $\rho_{\alpha,\Lambda}$ for the amplitude-damping channel is given by,
\begin{equation}
\label{eq-trho}
T_{\rho_{\alpha,\Lambda}}=\mbox{Diag}(\alpha,\alpha,1) \begin{pmatrix} R_{11} & R_{21} & R_{31} \\ R_{12} & R_{22} & R_{32} \\ R_{13} & R_{23} & R_{33} \end{pmatrix} \mbox{Diag}(\sqrt{(1-p)},-\sqrt{(1-p)},(1-p)) + 
\begin{pmatrix} 0 & 0 & 0 \\ 0 & 0 & 0 \\ 0 & 0 &  (2\lambda - 1)p \end{pmatrix}  ,
\end{equation}

with $\alpha=2\sqrt{\lambda(1-\lambda)}$ . 

Now consider $|\tilde{\phi}(\lambda) \rangle = \sqrt{\lambda} |00 \rangle + \sqrt{1- \lambda} |11 \rangle $ and let $\rho_1=\rho_{\tilde{\phi}(\lambda),\Lambda} $(eqn. (\ref{eqn1-rho1}). So we have a diagonal correlation matrix 
$T_{\rho_1}=(t_{ij})$ with $t_{11}= 2\sqrt{\lambda (1-\lambda)(1-p)}$ , $t_{22}= -2\sqrt{\lambda (1-\lambda)(1-p)}$ and
$t_{33}= \lambda + (1-\lambda)(1-2p)$ .Thus the condition  $M(\rho_1)=t_{11}^2 + t_{22}^2 \leq 1$ is satisfied. Also here,

$t_{11}^2 + t_{33}^2 = 4\lambda(1-\lambda)(1-p) + \lambda^2 + (1-\lambda)^2(1-2p)^2 + 2\lambda (1-\lambda)(1-2p)$, with the 
non-locality breaking condition $p\geq \frac{1}{2}$.  We have therefore
$t_{11}^2 + t_{33}^2 < 2\lambda(1-\lambda) + \lambda^2 + (1-\lambda)^2 = 1$ . Thus we see that if the 
amplitude damping channel $\Phi$ breaks the non-locality of the maximally entangled state $|\Phi^+\rangle$ , it then also breaks the non-locality  
in the states  $|\tilde{\phi}(\lambda) \rangle = \sqrt{\lambda}|00\rangle + \sqrt{(1-\lambda)}|11\rangle $ . 

The maximal Bell violation for action of the amplitude damping channel on all pure entangled states  is given by ,
\begin{equation}
\label{M-ampl}
M= \mbox{max}_{\{\lambda \in [0,1],R \in SO(3) \}}  \{ \sigma_1(t')^2  + \sigma_2(t')^2 \},
\end{equation}
where $ \sigma_1(t')$ and $\sigma_2(t')$ denote the first two singular values in descending order of $T_{\rho_{\alpha,\Lambda}}$ in eqn.(\ref{eq-trho}) .

\begin{figure}
\includegraphics[width=90mm]{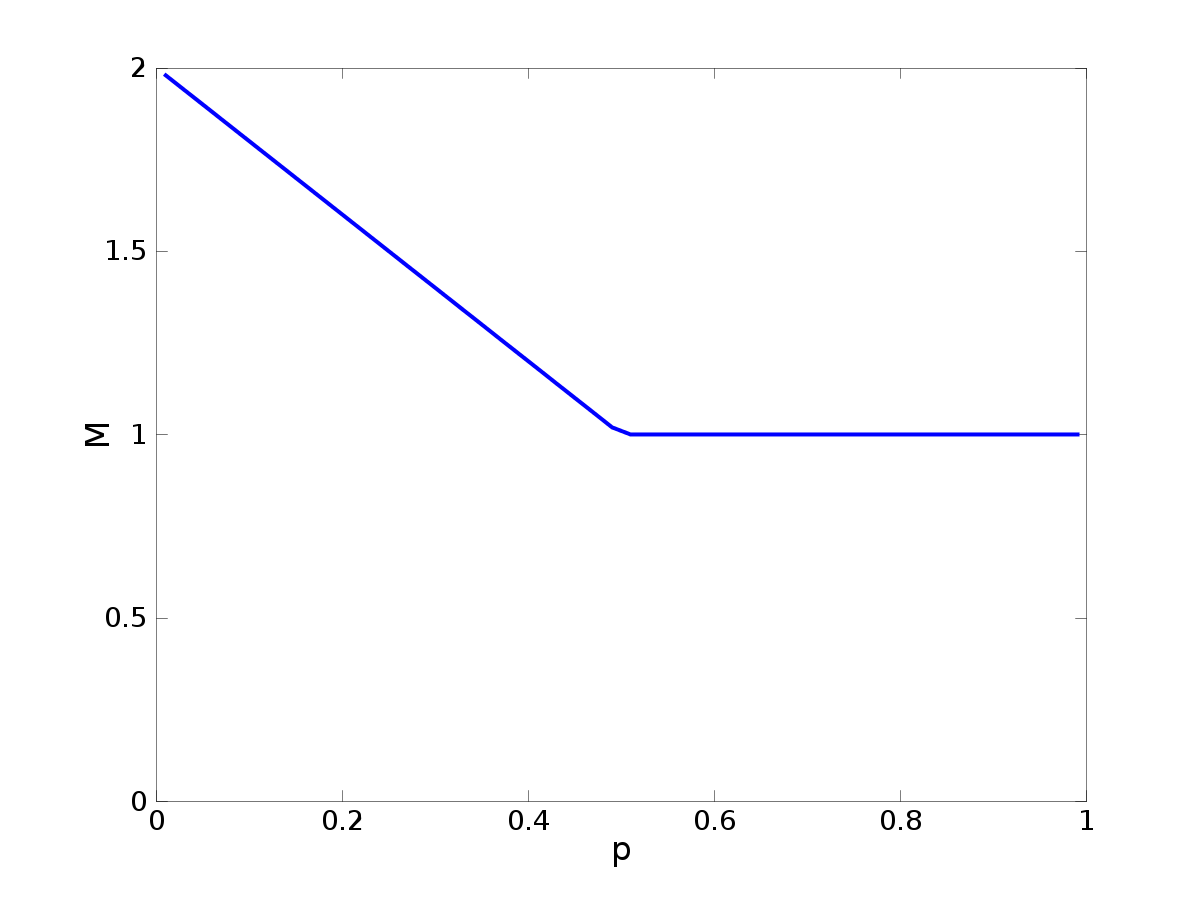}
\caption{p vrs. M }
\label{fig-ampldamp}
\end{figure}

\paragraph{Numerics:}
For the purpose of numerical investigation we take the same decomposition of $R$ as in eqn. (\ref{euler}). The maximization in eqn.(\ref{M-ampl}) has been done by choosing  $\alpha_0,\beta_0 , \gamma_0 \in \left[0,2\pi\right] $  with 
an interval of $0.1$ for each and
$\lambda \in \left[0,1\right]$ with an interval of $0.05$. 
Figure \ref{fig-ampldamp} shows the variation of M with respect to $p$ varying between $0$ and $1$ . $M$ is very close to $2(1-p)$ for $p \leq 0.5$ and exactly equals to 
$1$ for all points $p > 0.5$. Thus we see from fig. \ref{fig-ampldamp} that the amplitude damping channel breaks non-locality of every state for $p \geq \frac{1}{2}$ .  
 
But as we will see in  section V it fails to be strongly non-locality breaking for any value of $p$. 

\subsection{Channels breaking non-locality of maximally entangled states also break that of states whose free sided reduction is maximally mixed}
It is however true that  if for a qubit channel $\$_B$ , $(I_A \otimes \$_B) (|\Phi^+\rangle \langle \Phi^+|)$ is a local state ({\it{i.e}}, eqn. (\ref{nonloc-condition}) 
is satisfied) , then for any two-qubit state $\sigma_{AB}$ , $(I_A \otimes \$_B)(\sigma_{AB})$
is also a local state provided $Tr_B(\sigma_{AB})= \frac{I_2}{2}$, $I_2$ being the $2 \times2$ identity matrix. This is proved in the Appendix.

In our aforesaid proof we see that $(I_A \otimes \$_B)((I_A \otimes \$_1)(|\Phi^+\rangle \langle\Phi^+|))$ 
is a local state if $(I_A \otimes \$_B)(|\Phi^+ \rangle \langle \Phi^+|)$ is a local state. From the structure of the proof (see Appendix) it is also clear that  
$(I_A \otimes \$_1)((I_A \otimes \$_B)(|\Phi^+\rangle \langle\Phi^+|))$ is also a local state  (we have to choose $A=R_W{D_1'}^2R_W^T$ , $B=D_1^2$ and use lemma 1). Thus we see that 
the composition of a qubit channel , that breaks non-locality of a maximally entangled state, with any other qubit channel
also does the same job.

\section{Stronger non-locality breaking}
In this section we consider strongly non-locality breaking channels.As proved in Lemma 3 of Section III, 
 for a channel $\Lambda$ to be strongly non-locality breaking  it is enough for the Choi-state of the channel $\rho_{\Phi^+,\Lambda}$ to not show any hidden non-locality.      

Hence, building on the work done by Verstraete et. al  in \cite{VW02},  \cite{Ver} and \cite{Ver-lor} we first provide a necessary and sufficient condition for an 
arbitrary two-qubit state to show hidden CHSH non-locality (\ref{hidden-nonlocality-subsection}). This is proved in Theorem 1.

Under a local filtering transformation taking any two-qubit state $\rho$ to \begin{equation}\rho'=(A \otimes B)\rho (A^{\dagger} \otimes B^{\dagger}), \label{local-filtering-eq} \end{equation} 
the real $4 \times 4$ matrix $R$ with $R_{ij}=Tr(\rho \sigma_i \otimes \sigma_j), i,j=0,1,2,3$ (where $\sigma_0=I_2 $) transforms as 
\begin{equation}
\label{snlb-1}
R'\equiv Tr(\rho'\sigma_i \otimes \sigma_j)=L_A R L_B^T |det(A)||det(B)|
\end{equation}
with the Lorentz 
transformations $L_A$ and $L_B$ being given by,
\begin{eqnarray}
L_A= \frac{T(A \otimes A^*)T^{\dagger}}{|det(A)|}, \nonumber \\ 
L_B=  \frac{T(B \otimes B^*)T^{\dagger}}{|det(B)|},
\end{eqnarray}
and $ T= \frac{1}{\sqrt{2}} \begin{bmatrix} 1 & 0 & 0 & 1 \\ 0 & 1 & 1 & 0 \\ 0 & i & -i & 0 \\ 1 & 0 & 0 & -1  \end{bmatrix}$, with the  normalisation factor   $R'_{00}=Tr(\rho')= Tr(A^{\dagger}A 
\otimes B^{\dagger}B\rho)$. 

{\textbf{Remark}}: The filters should be of full rank, for the Lorentz transformations to be finite. 

Further, let $C_{\rho}=MRMR^T$ with  $M=\mbox{diag}(1,-1,-1,-1)$.

\vspace{10mm}
{\textbf{Theorem 1:}} Let $\lambda_i(C_{\rho}),(i=0,1,2,3)$ denote the eigenvalues of $C_{\rho}$ in descending order for an arbitrary two-qubit  state $\rho$. Then, $\rho$ shows hidden 
CHSH nonlocality iff 
\begin{equation}
\lambda_1(C_{\rho}) + \lambda_2(C_{\rho}) > \lambda_0(C_{\rho}). 
\label{hid-nlc-cond}
\end{equation}
The maximum  Bell violation obtained from the   optimal filtered (or quasi-distilled) Bell-diagonal state being $2\sqrt{\frac{(\lambda_1(C_{\rho}) + \lambda_2(C_{\rho}))}{\lambda_0(C_{\rho})}}$. 

\vspace{10mm}
{\textbf{Proof}}:

It was  shown in \cite{Ver} and \cite{Ver-lor} that by suitably choosing $A$ and $B$ and hence proper orthochronous Lorentz transformations $L_A$ , $L_B$
for {\textit{any}} $\rho$ we can have $R'$ to be either diagonal corresponding to a Bell-diagonal state $\rho'$ or of the  form,
\begin{equation}
 R'=R_{\rho'} = \begin{bmatrix} a & 0 & 0 & b \\ 0 & d & 0 & 0 \\ 0 & 0 & d & 0 \\ c & 0 & 0 & (b+c-a)   \end{bmatrix} 
\end{equation}
with   the corresponding $\rho'$(unnormalized) being 
\begin{equation}
\label{rho-after-local-filtering}
\rho'= \frac{1}{2 }\begin{bmatrix} b+c & 0 & 0 & 0 \\ 0 & a-b & d & 0 \\ 0 & d & (a-c) & 0 \\ 0 & 0 & 0 &  0   \end{bmatrix} .
\end{equation}
The  possible sets of real values of $b$, $c$ and $d$ are given by,
\begin{eqnarray}
&({\rm i})& b=c=\frac{a}{2}, \nonumber \\ 
&({\rm ii})& (d=0=c) \mbox{  and  }  (b=a),  \nonumber \\ 
&({\rm iii})& (d=0=b) \mbox{  and  } (c=a), \nonumber \\ 
&({\rm iv})& (d=0) \mbox{ and } (a=b=c) .  \label{value-cases}
\end{eqnarray}
Case (i) corrsponds to rank three or two states while the other cases corrrespond to either the product states $|00\rangle \langle 00|$ or the state $|0\rangle \langle 0| \otimes \frac{I}{2}$. 

\vspace{5mm}
From eqn. (\ref{snlb-1}) it follows that  the spectrum of $MR'MR'^T$ is given by
\begin{equation}
\label{c-matrix-transformation}
\lambda(MR'MR'^T) = |det(A)|^2  |det(B)|^2 \lambda(ML_ARL_B^TML_BR^TL_A^T) =|det(A)|^2  |det(B)|^2 \lambda(MRMR^T),   
\end{equation}
where we have used $L_A^TML_A=M=L_B^TML_B$. 
Now as the filters $A$ and $B$ are of full rank i.e, $det(A),det(B) \ne 0$ we have for each for each $i \in \{0,1,2,3\}$ 
\begin{equation}
\frac{\lambda_i({C_{\rho'}})}{\lambda_0(C_{\rho'})} =  \frac{\lambda_i({C_{\rho}})}{\lambda_0(C_{\rho})}. 
\label{eigfrac}
\end{equation}
%as $det(A),det(B)$ are both non-zero for proper orthochronous Lorentz transformations corresponding to full rank filters. 

Let us consider the following cases now.

\vspace{5mm}
(a)  $R'=\mbox{diag}(s_0,s_1,s_2,s_3)$.  $\rho'$  corresponds to a Bell-diagonal state which in turn violates the Bell-CHSH inequality (\cite{Hor95}) after normalization provided,
\begin{equation}
1< \frac{s_1^2}{s_0^2} + \frac{s_2^2}{s_0^2}= \frac{\lambda_1(C_{\rho'})
+ \lambda_2(C_{\rho'})}{\lambda_0(C_{\rho'})}=\frac{\lambda_1(C_{\rho})
+ \lambda_2(C_{\rho})}{\lambda_0(C_{\rho})}  \nonumber.
\end{equation}
(by eqn. (\ref{eigfrac})). This proves Thoerem 1 for this case.

\vspace{5mm}
(b) 
$\rho'$ is of the non Bell-diagonal form with $d \ne 0$ in eqn. (\ref{rho-after-local-filtering})  (case (i) of eqn. (\ref{value-cases})) .

\vspace{5mm}
It is easy to see by partial transposition that $\rho'$ must be entangled .

Further by using filters of the form of  
$A=\mbox{diag}(\sqrt{\frac{(a-c)}{(a-b)}}\frac{1}{n},1)$ and $B=\mbox{diag}(\frac{1}{n},1)$
we have, 
\begin{eqnarray}
\rho_1 &=& (A \otimes B) \rho (A^{\dagger} \otimes B^{\dagger}) \nonumber \\ 
&=& \frac{1}{2} \left(\frac{(b+c)(a-c)}{(a-b)n^4}| 00 \rangle \langle 00| + \frac{(a-c)}{n^2} (|01 \rangle \langle 01| + |10 \rangle \langle 10|) + 
\frac{d\sqrt{(a-c)}}{n^2\sqrt{(a-b)}} (|01 \rangle \langle 10| + |10 \rangle \langle 01|) \right) . 
\end{eqnarray}

By taking a very large positive no. $n$ , $\rho_2 = \frac{\rho_1}{Tr(\rho_1)}$ can be made to approach arbitrarily close to the Bell-diagonal state 
\begin{eqnarray}
\rho_3 &=& \frac{1}{2} ((|01 \rangle \langle 01| + |10 \rangle \langle 10|) + \frac{d}{\sqrt{(a-b)(a-c)}}(|01 \rangle \langle 10| + |10 \rangle \langle 01|)) \nonumber \\
&=& \frac{1}{4} ( I \otimes I + \frac{d}{\sqrt{(a-c)(a-b)}} \sigma_1 \otimes \sigma_1 + \frac{d}{\sqrt{(a-c)(a-b)}} \sigma_2 \otimes \sigma_2 - \sigma_3 \otimes \sigma_3 ).
\end{eqnarray}
Now,  from eqn. (\ref{rho-after-local-filtering})  we have $\lambda(C_{\rho'})= [(a-b)(a-c), (a-b)(a-c), d^2, d^2 ]$. 

From theorem 3 of ref. \cite{VW02} we also know that the optimal Bell-violation among the states connected to $\rho$ by local filtering transformations is obtained from the `quasi-distilled' state $\rho_3$. Hence by using eqn. (\ref{eigfrac}) 
we obtain an optimal Bell violation of amount

\begin{equation}
1 + \frac{d^2}{(a-b)(a-c)} =  \frac{\lambda_1(C_{\rho'})
+ \lambda_2(C_{\rho'})}{\lambda_0(C_{\rho'})} = \frac{\lambda_1(C_{\rho})
+ \lambda_2(C_{\rho})}{\lambda_0(C_{\rho})} > 1 
\end{equation}
(note that $(a-b)(a-c) \geq d^2$ by virtue of positivity of $\rho'$)

Thus states for which $\rho'$ is not Bell-diagonal ($d \ne 0$ case ) will {\textit{always}} violate the Bell-CHSH inequality after suitable local filtering transformation. 

\vspace{5mm}
(c)
$\rho'$ is of the non Bell-diagonal form with $d = 0$ in eqn. (\ref{rho-after-local-filtering}) (cases (ii), (iii) and (iv) of eqn. (\ref{value-cases})) . 
These states being of the product form must come from a separable $\rho$ (local filtering  with full rank filters being  invertible) and from eqns. (\ref{rho-after-local-filtering}) and (\ref{c-matrix-transformation}) we have $\lambda_i(C_{\rho})=\lambda_i(C_{\rho'})=0$ for all $i$. Thus Theorem 1 holds.  

\vspace{3mm}
Conversely, when eqn. (\ref{hid-nlc-cond}) is satisfied we can either filter or quasi-distill $\rho$ to a Bell-diagonal state with optimal Bell-violation $2\sqrt{\frac{(\lambda_1(C_{\rho}) + \lambda_2(C_{\rho}))}{\lambda_0(C_{\rho})}}$.

\hfill $\square$ 

\subsection{Strongly non-locality breaking qubit channels}
Theorem 1 allows us to characterise    
exactly all the strongly non-locality breaking qubit channels within the purview of CHSH nonlocality.

 Let us first consider the amplitude damping channel which  breaks the non-locality  
of all states for $p \geq \frac{1}{2}$(see fig. \ref{fig-ampldamp}) . The Choi-state of the amplitude damping channel is given by,

\begin{equation}
\rho_1= \frac{1}{2} \begin{bmatrix} 1 & 0 & 0 & \sqrt{(1-p)} \\ 0 & 0 & 0 & 0 \\ 0 & 0 & p & 0 \\ \sqrt{(1-p)} & 0 & 0 & (1-p)  \end{bmatrix} .
\end{equation}
The eigenvalues of $C_{\rho_1}$ are given by, $[(1-p),(1-p),(1-p),(1-p)] $ and eqn. (\ref{hid-nlc-cond}) is satisfied provided $p < 1$. As can be checked, the Choi-state can be quasi-distilled to a singlet state using the filters 
$A=\mbox{diag}(\frac{(1-p}{(2-p)},\frac{1}{n})$ and $B=\mbox{diag}(\frac{1}{n},1)$. Hence, a channel of the amplitude damping form can never be strongly non-locality
breaking for any non-zero value of  $p$ even though it breaks the nonlocality of all states for $p \geq \frac{1}{2}$. Note that amplitude damping channels are also not
entanglement breaking for any non-zero $p$.  

\subsubsection{Unital channels}
For unital qubit channels we have $\vec{t}=0$ in eqn. (\ref{map-canonical}) and the Choi state $\rho_{\Phi^+,\Lambda}$ is local unitarily connected to the Bell-diagonal state,
\begin{equation}
\rho_{\Phi^+, \Lambda'}= \frac{1}{4}( I \otimes I +  \lambda_1 \sigma_1 \otimes \sigma_1 - \lambda_2 \sigma_2 \otimes \sigma_2 +  \lambda_3 \sigma_3 \otimes \sigma_3) .
\end{equation} 
As this itself is of the normal form, from Theorem 1 and eqn .(\ref{nlb-mes-1}) it follows that   unital channels breaking non-locality of maximally 
entangles states are also strongly non-locality breaking. Hence, as mentioned after eqn.(\ref{choi-state-filter}), they also break the non-locality of any input state. 

\subsubsection{Extremal qubit channels}
The set of all qubit channels is convex and it was shown in \cite{MB02} that the closure of the set of extreme points of this set  are given, upto pre and post-processing by  unitaries, by a two parameter 
family with the canonical form (eqn. (\ref{map-canonical})), 
\begin{equation}
 {\mathbf{T}}_{\Lambda} = 
\begin{pmatrix} 1 & 0 & 0 & 0 \\ 0 & cos(u) & 0 & 0 \\ 0 & 0 & cos(v) & 0 \\ sin(u)sin(v) & 0 & 0 & cos(u)cos(v)  \end{pmatrix} ,
\end{equation}
with $u \in [0,2\pi), v \in [0,\pi)$ . 
The $R$ matrix of Choi-state of this extremal channel is given by,
\begin{equation}
\begin{pmatrix} 1 & 0 & 0 & 0 \\ 0 & cos(u) & 0 & 0 \\ 0 & 0 & -cos(v) & 0 \\ sin(u)sin(v) & 0 & 0 & cos(u)cos(v)  \end{pmatrix} ,
\end{equation},
with $\lambda(C_{\rho_{\Phi^{+},\Lambda}})=[cos^2(u),cos^2(u),cos^2(v),cos^2(v)]$ assuming $cos(u) > cos(v)$. The $M-$value for the optimal Bell-violation (eqn.
\ref{nonloc-condition}) of the filtered Bell-diagonal version of $\rho_{\Phi^+,\Lambda}$, from Theorem 1 is given by $\left(1 + \left({\frac{cos(v)}{cos(u)}}\right)^2 > 1\right)$ and hence an extremal channel cannot be strongly 
non-locality breaking. As a check one can compute the eigenvectors of $C_{\rho_{\Phi^{+},\Lambda}}$ and $C'_{\rho_{\Phi^{+},\Lambda}}$ to obtain the filters(see 
\cite{VW02} for details) 
$A=i \begin{pmatrix} 0 & {(\frac{sin(\frac{(v+u)}{2})cos(\frac{(v-u)}{2})}{sin(\frac{(v-u)}{2})cos(\frac{(v+u)}{2})})}^{\frac{1}{4}} \\   
{(\frac{sin(\frac{(v-u)}{2})cos(\frac{(v+u)}{2})}{sin(\frac{(v+u)}{2})cos(\frac{(v-u)}{2})})}^{\frac{1}{4}}            &    0             \end{pmatrix}$

and

$B= \begin{pmatrix}  {(\frac{sin(\frac{(v+u)}{2})cos(\frac{(v+u)}{2})}{sin(\frac{(v-u)}{2})cos(\frac{(v-u)}{2})})}^{\frac{1}{4}}  & 0 \\               
0 & {(\frac{sin(\frac{(v-u)}{2})cos(\frac{(v-u)}{2})}{sin(\frac{(v+u)}{2})cos(\frac{(v+u)}{2})})}^{\frac{1}{4}}                        \end{pmatrix},$

through which  the Choi-state of extremal channels can be brought to the Bell-diagonal state ,
\begin{equation}
\rho_3 = \frac{1}{4} (  I \otimes I +  \sigma_1 \otimes \sigma_1 + \frac{cos(v)}{cos(u)} \sigma_2 \otimes \sigma_2 - \frac{cos(v)}{cos(u)} \sigma_3 \otimes \sigma_3  ).
\end{equation}

If $cos(u)$ or $cos(v)=0$ then the channel becomes entanglement breaking and hence also strongly non-locality breaking. 
The condition for an extremal channel to break 
nonlocality of a maximally entangled state is of course  $cos^2(u) + cos^2(v) \leq 1$.

\subsubsection{Channels breaking nonlocality of maximally entangled state genuinely, may not be strongly non-locality breaking }
In ref. \cite{Gen-hid-quan-nlc} an example was given of a one parameter family of two-qubit entangled states which have a local model 
for projective measurements but shows hidden CHSH nonlocality under suitable filtering. The one parameter family 
of states, defined by the parameter $q$, is 
given by  $\rho=qP\{\frac{1}{\sqrt{2}}(|01\rangle - |10\rangle)\} + (1-q)|0 \rangle \langle 0| \otimes \frac{I}{2} $ where 
$P\{|\alpha\rangle\}$ denotes projector on $|\alpha\rangle$. This state is dual to the channel $\$$ with $\vec{t}=(0,0,(1-q))$ and 
$\vec{\lambda}=(-q,q,-q)$ (see eqn. (\ref{map-canonical})), i.e., $\rho = (\$ \otimes I)(|\Phi^+ \rangle \langle \Phi^+|)$. The state is entangled for all values of $q >0$ and has a local model for projective measurements for all $q \leq \frac{1}{2}$. 
Hence $\$$ is not entangled breaking for  
any positive value of $q$ and breaks the non-locality of maximally entangled states genuinely
(in the sense that the output has a local model for projective measurement ) for $q \leq \frac{1}{2}$. It turns out that
 $\lambda(C_{\rho})=[q,q,q^2,q^2]$.
 Thus, by Theorem 1 and as mentioned in ref. \cite{Gen-hid-quan-nlc}, the optimal Bell-violation under local filtering is $2\sqrt{1 + q}$ 
and the channel is not strongly nonlocality breaking for any positive value of $q$. 
 The channel breaks CHSH non-locality of maximally entangled states for $q \leq \frac{1}{\sqrt{2}}$. Interestingly, we find numerically  that $\$$  {\textit{fails}} to  break the CHSH non-locality  of an arbitrary 
input state for the range $ 0.62 < q  \leq \frac{1}{\sqrt{2}} $ (the lower bound is correct upto two decimal places, for $q=0.6236$ M-value of  the state (eqn. (\ref{M-value})) $(I \otimes \$)(\sqrt{\lambda}|00\rangle + \sqrt{1-\lambda}|11\rangle)$ for $\lambda=0.95$ being 
$1.0+ 2.339 \times 10^{-5}$). Thus the question of whether  the channel genuinely breaks the nonlocality of all states for $q\leq \frac{1}{2}$ remains open .

\subsubsection{Example of non-unital strongly non-locality breaking channels}
Using Theorem 1 one can easily generate examples of non-unital strongly non-locality breaking channels. For example the channel in  canonical form (\ref{map-canonical}) with parameters 
 $t_1=0, t_2=0,t_3=0.29, \lambda_1=\frac{1}{\sqrt{2}}, \lambda_2=\frac{1}{\sqrt{10}}, \lambda_3=\frac{1}{2} $ has for $\rho \equiv \rho_{\Phi^+,\Lambda}$, $\frac{\lambda_1(C_{\rho}) + \lambda_2(C_{\rho})}{\lambda_0(C_{\rho})} = 
 0.887$ and hence is strongly non-locality breaking by Theorem 1.

\subsection{Relative vol. of strongly non-locality breaking channels  and  entanglement breaking channels}
Entanglement breaking channels are isomorphic to the set of separable states whose one-sided reduction is maximally mixed. As Lemma 3 shows , through the Choi-Jamiolkowski isomorphism strongly non-locality breaking qubit   
channels are isomorphic to the set of  states which do not show any hidden nonlocality, with one sided reduction maximally mixed. As a quantitative comparison of entanglement and non-locality it thus becomes interesting to compute the volume of this set and 
Theorem 1 allows us to achieve this.   
So, we compare  the relative volume(w.r.t the volume of all qubit channels) of the set of all strongly non-locality breaking qubit channels(also a convex set) with that of the entanglement breaking channels. For this we sample uniformly
within a six dimensional real hypercube of parameters $t_i, \lambda_i \in [-1,1], i=1,2,3$ and reject points \footnote{Complete positivity of qubit channels with the canonical form given by eqn.(\ref{map-canonical})  
demands that $|t_i|,|\lambda_i| \leq 1$ \cite{MB02}. }  which do not satisfy complete-positivity criterion. Among the remaining       
points we count the fraction which correspond to strongly non-locality breaking channels using theorem 1 and the fraction that correspond to entanglement breaking channels. We also count the 
fraction which breaks just the non-locality of maximally entangled states. We sample $10^7$ points for this purpose. The relative volume of the entanglement breaking channels turn out to be about 0.24, that of 
channels breaking non-locality of maximally entangled state turn out to be about 0.81 , while that of strongly non-locality breaking channels turn out to be about 0.39.
 If we restrict to the unital case then the vol. of entanglement breaking channels is   0.5 and that of non-locality breaking channels turns out to be 0.92. Thus though almost any unital channel is strongly non-locality breaking the chance 
that a generic qubit channel is strongly non-locality breaking is much closer to that of it being entanglement breaking.

\section{Discussion}
In this work inspired by the notion of entanglement breaking we investigate qubit channels through their property of `non-locality breaking'. This additionally,provides a way  to connect the notion of entanglement and non-locality through channel action---instead of the usual
trend via states.
 We focus on CHSH nonlocality as this is the only inequality for which the necessary and sufficient conditions  
on the state for violation are known. One of the main properties of entanglement breaking channels is that it is sufficient to `break' the entanglement of maximally entangled states.  We provide examples to show that similar property does not hold for `non-locality breaking'.Though there seems to be some channels and a certain restricted class of states for all channels for which this is true.
We also consider a stronger notion of non-locality breaking, again taking cue from entanglement breaking where the output states of one-sided action of the channel are required to be local under SLOCC. 
We show that  for a qubit channel to be strongly non-locality breaking it is enough for the dual-state of the channel to not show any hidden nonlocality under local filtering. 

%We show that within the purview of CHSH nonlocality, the aforesaid isomorphism is restored in the sense that for a qubit channel to be strongly non-locality breaking 
%it is enough for the dual-state of the channel to not show any hidden nonlocality under local filtering. 

We  provide a closed-form necessary sufficient condition for any two-qubit state to violate the Bell-CHSH inequality under local filtering,which is likely to be useful 
for other purposes as well. This is then used to study `strongly non-locality breaking' qubit channels and compute their relative volume within the set of all channels. It turns out that unital qubit channels breaking non-locality of maximally entangled states are strongly non-locality breaking while extremal
qubit channels cannot be so unless they are entanglement breaking. It may be mentioned here that each single mode entanglement breaking Gaussian channel is related to a single mode non-classicality breaking Gaussian channel  
via some squeezing transformation \cite{sol-03}.

An interesting course of future study is to see how the gap between entanglement breaking and non-locality breaking qubit channels close as one considers  more inequalities ( e.g,$I_{3322}$ \cite{Collins-Gisin-2003}). We also, at present, do not have any  
example of a channel which breaks the non-locality of a maximally entangled state, {\textit{genuinely}} but fails to break that of other states. This can be studied, for example, for  the channel in Section V whose dual state has a local model.

\section{Appendix}
 
 \paragraph{Lemma 4:} Suppose $A$ and $B$ are positive definite $3\times3$ matrices. Let $\lambda_i(A)$ denote the {\it{i}}-th eigenvalue of $A$ in the descending order and $\lambda_i(B)$ be that of $B$ . Also let $\lambda_i(B) \leq 1$ and 
$\lambda_1(A) + \lambda_2(A) \leq 1$. If $\Lambda_i$  denote the \textit{i}-th eigenvalue of $AB$ in descending order then we have,  
\begin{equation}
\label{lemma}
\Lambda_1 + \Lambda_2 \leq 1  .
\end{equation}
Proof :
According to  Marshall
and Olkin \cite{MO79},  for any $n\times n$ real matrices $A,B$ : 
\begin{equation}
\label{olkin}
-\sum_{i=1}^n \sigma_{[i]}(A)  \sigma_{[i]}(B) \leq  Tr(AB)  \leq   \sum_{i=1}^n \sigma_{[i]}(A)  \sigma_{[i]}(B)
\end{equation}

where $\sigma_{[i]}(A)$ and $\sigma_{[i]}(B)$ denote the \textit{i}-th singular values of $A$ and $B$ respectively. Thus, applying eqn. (\ref{olkin}) we have,
\begin{equation}
\label{Lambda}
\Lambda_1 + \Lambda_2 + \Lambda_3 \leq \lambda_1(A) \lambda_1(B) +  \lambda_2(A) \lambda_2(B) + \lambda_3(A) \lambda_3(B) .
\end{equation}

Again, corollary 2.4 of \cite{LU00} says that if $A$ and $B$ are positive definite Hermitian matrices, then for each $k=1,2,...n$,
\begin{equation}
min \{\lambda_1(A) \lambda_k(B) ,  \lambda_1(B) \lambda_k(A) \} \geq \lambda_k(BA) \geq max \{ \lambda_n(A)\lambda_k(B), \lambda_n(B)\lambda_k(A) \}
\end{equation}
where the eigenvalues $\lambda_1,\lambda_2, ...,\lambda_n$ of each matrix are arranged in the descending order. Hence it follows that,
\begin{equation}
\label{lambda3-bound}
\Lambda_3 - \lambda_3(A)\lambda_3(B) \geq 0 .
\end{equation}

Thus  from eqn. (\ref{Lambda}) and eqn. (\ref{lambda3-bound}) it follows that,
\begin{eqnarray}
\Lambda_1 + \Lambda_2  \leq \lambda_1(A) \lambda_1(B) +  \lambda_2(A) \lambda_2(B) \leq  \lambda_1(A)  +  \lambda_2(A)  (\mbox{  as   } \lambda_i(B) \leq 1 )  \nonumber \\  
 \leq  1 .
\end{eqnarray}

\hfill $\square$ . 
 
\subsection{Proof of the proposition  ``Channels breaking non-locality of maximally entangled states also break that of states whose free sided reduction is maximally mixed''}
At par with the notations used in subsection IV. C,
from eqn. (\ref{map-composition}) it follows  that $\$_B$ can always be expressed as  $\$_B= U_B \circ \$'_B \circ V_B $ , where $U_B$ and $V_B$ are 
unitary channels while $\$'_B$ is  a qubit channel in the canonical form of eqn. (\ref{map-canonical}) . Now,
$(I_A \otimes \$_B) (|\Phi^+\rangle \langle \Phi^+|)=(I \otimes U_B)((I \otimes \$_B') ((I \otimes V_B)|\Phi^+ \rangle \langle \Phi^+ |(I_A \otimes V_B^{\dagger}))
(I \otimes U_B^{\dagger}) = 
(V_B^T \otimes U_B) ((I_A \otimes \$_B') (|\Phi^+ \rangle \langle \Phi^+ |))(V_B^* \otimes U_B^{\dagger})$ . 
So,  $(I_A \otimes \$_B') |\Phi^+ \rangle \langle \Phi^+ |$ is also a local state if $ (I_A \otimes \$_B) |\Phi^+ \rangle \langle \Phi^+ |$ is local.

As $\sigma_{AB}$ is a two-qubit density matrix with $Tr_B(\sigma_{AB})=\frac{I_2}{2}$, therefore by the Choi-Jamiolkowski isomorphism (\cite{JAM72}) , 
there exists a trace preserving qubit channel $\$_1$ such that $\sigma_{AB}=(I \otimes \$_1 )(|\Phi^+ \rangle \langle \Phi^+ |)$. 
Once again we can represent $\$_1$ as: $\$_1=U_1 \circ \$'_1 \circ V_1$ with $U_1$ and $V_1$ being unitary channels while $\$'_1$ is in 
the canonical form (\ref{map-canonical}). So here,

\begin{eqnarray}
 (I_A \otimes \$_B) (\sigma_{AB}) &=&  (I_A \otimes \$_B \circ  \$_1) (|\Phi^+\rangle \langle \Phi^+|) \nonumber\\
&=& (I_A \otimes  U_B \circ \$'_B \circ V_B \circ  U_1 \circ \$'_1 \circ V_1) (|\Phi^+\rangle \langle \Phi^+|)  \nonumber\\ 
&=& (V_1^T \otimes U_B) ((I_A \otimes \$'_B \circ  V_B  U_1 \circ \$'_1)  (|\Phi^+\rangle \langle \Phi^+|)) (V_1^* \otimes U_B^{\dagger}) .
\end{eqnarray}
 
Thus to check whether the state $(I_A \otimes \$_B) (\sigma_{AB})$ is local, it is enough to check whether the state 
\begin{equation}
\label{eq-rho2}
\rho_2 \equiv (I_A \otimes \$'_B \circ  V_B  U_1 \circ \$'_1)  (|\Phi^+\rangle \langle \Phi^+|) 
\end{equation}

 is local.   Now the $\mathbf{T}$ matrix 
corresponding to the channel $\$'_{B} \circ W \circ \$'_{1} $ (considering the matrix $V_BU_1$ as $W$) is given by:  

\begin{equation}
\label{eq-tproduct}
\mathbf{T}_{\$'_{B} \circ W \circ \$'_{1}} = \begin{pmatrix}
1 & 0 & 0 & 0 \\ 
t'_1 & \lambda'_1 & 0 & 0  \\ 
t'_2 & 0 & \lambda'_2 & 0 \\
t'_3 & 0 & 0 & \lambda'_3 
\end{pmatrix}
\begin{pmatrix}
1 & 0 & 0 & 0 \\ 
0 & w_{11} & w_{12} & w_{13} \\
0 &  w_{21} & w_{22} & w_{23}  \\
0 &  w_{31} & w_{32} & w_{33} 
\end{pmatrix}
\begin{pmatrix}
1 & 0 & 0 & 0 \\ 
t_1 & \lambda_1 & 0 & 0 \\
t_2 & 0 & \lambda_2 & 0 \\
t_3 & 0 & 0 & \lambda_3
\end{pmatrix},  
\end{equation}
with the $\mathbf{T}$ matrices for  $\$'_B$ , $W$ and  $\$'_1$ being the first , second and the third $4 \times 4$ matrix on the RHS of eqn. (\ref{eq-tproduct})
from left to right. Note that here the $3 \times3$ rotation matrix corresponding to 
$W \in SU(2)$ is given by,
\begin{equation}
R_W = \begin{pmatrix}
       w_{11} & w_{12} & w_{13} \\
       w_{21} & w_{22} & w_{23} \\
        w_{31} & w_{32} & w_{33} \\
      \end{pmatrix}
 \equiv \begin{pmatrix} \vec{w_1} \\ \vec{w_2} \\ \vec{w_3} \end{pmatrix} (\mbox{say}) .
\end{equation}

So from eqn. (\ref{eq-tproduct}) , we get:
\begin{equation}
\mathbf{T}_{\$'_{B} \circ W \circ \$'_{1}} =  \begin{pmatrix}
1 & 0 & 0 & 0 \\ 
t'_1 + \lambda_1'(\vec{w_1}.\vec{t})&   &   &   \\
t'_2 + \lambda_2'(\vec{w_2}.\vec{t}) &  & D_1'R_WD_1  &  \\
t'_3 + \lambda_3'(\vec{w_3}.\vec{t}) &  &  & 
\end{pmatrix}  
\end{equation}

where $\vec{t}= \begin{pmatrix} t_1 \\ t_2 \\ t_3 \end{pmatrix}$ , $D_1= diag(\lambda_1,\lambda_2,\lambda_3)$ and $D'_1=diag(\lambda_1',\lambda_2',\lambda_3')$ . Let the singular value decomposition of $D'_1R_WD_1$ be $R_1DR_2$ ,
where $R_1$ and $R_2$ are $3 \times3$ real rotation matrices and D is a $3 \times 3$ real diagonal matrix consisting of the singular values of $D_1'WD_1$. Then we have,

\begin{equation}
\mathbf{T}_{\$'_{B} \circ W \circ \$'_{1}}  =  \mathbf{T}_{W_1 \circ \$_{\alpha} \circ W_2} ,              \nonumber 
\end{equation}
with $W_j$ being the unitary map corresponding to the rotation $R_j$ and $\$_{\alpha}$ is a map for which ,
\begin{equation}
\mathbf{T}_{\$_{\alpha}} = \begin{pmatrix}  
1 & 0 & 0 & 0 \\ 
t''_1 &   &   &   \\
t''_2  &  & D  &  \\ 
t''_3  &  &  & 
\end{pmatrix}  .
\end{equation}
Here, from eqn. (\ref{eq-rho2}),
\begin{equation}
\rho_2= (I_A \otimes W_1) ((I_A \otimes \$_{\alpha})((I_A \otimes W_2 )|\Phi^+ \rangle \langle \Phi^+|(I_A \otimes W_2^{\dagger} ) ))  (I_A \otimes W_1^{\dagger})= 
(W_2^T \otimes W_1) 
((I_A \otimes \$_{\alpha})(|\Phi^+\rangle \langle\Phi^+|)) (W_2^* \otimes W_1^{\dagger}) .
\end{equation}

So, in order to show that $\rho_2$ is a local state , it is enough to show that $\rho_3= (I_A \otimes \$_{\alpha})(|\Phi^+\rangle \langle\Phi^+|) $  is a local state, i.e it satisfies eqn. (\ref{nonloc-condition}). As $\$_{\alpha}$ is a valid
channel , therefore, in accordance with eqn. (\ref{nlb-mes-1}) , the condition for the state $\rho_3$ being local is given by,   
\begin{equation}
\label{rho3local}
\Lambda_1 + \Lambda_2 \leq 1 , 
\end{equation}
where $\Lambda_1$ and  $\Lambda_2$ denote the squares of the two larger eigenvalues of the $3 \times 3$ diagonal matrix D. Now, by the construction of D , $\Lambda_1$ and $\Lambda_2$ are also the two larger eigenvalues of  
$(D_1'R_WD_1)(D_1'R_WD_1)^T = D_1'R_WD_1^2R_W^TD_1'$. And so, $\Lambda_1$ , $\Lambda_2$ are the two larger eigenvalues of  $D_1'^2R_WD_1^2R_W^T$ \cite{HJ90} .

Let us now choose $A={D_1'}^2$ and $B=R_WD_1^2R_W^T$. As the channel $\$_B$ (and therefore, the channel $\$_B'$) is assumed to be non-locality breaking , therefore , by eqn. (\ref{nonloc-condition}) , we must have: ${\lambda'_1}^2 + {\lambda'_2}^2 \leq 1$ with 
$|\lambda'_3| \leq min \{ |\lambda'_1|,|\lambda'_2| \}$ . Thus here $A$ is a positive semi-definite $3 \times 3$ matrix with its eigenvalues being ${\lambda'_1}^2$ ,  ${\lambda'_2}^2$ ,  ${\lambda'_3}^2$  in descending order such that
${\lambda'_1}^2 + {\lambda'_2}^2 \leq 1 $ . 

On the other hand , $\$_1$ (and there by $\$_1'$) being a quantum channel , without loss of generality , we can take in eqn. (\ref{eq-tproduct}) that $\lambda_1^2$,   $\lambda_2^2$ and   $\lambda_3^2$ (the eigenvalues of B) are in descending order and
$\lambda_i^2 \leq 1$ for $i=1,2,3$.

Then by lemma 4, $\Lambda_1 + \Lambda_2 \leq 1$ , which is nothing but eqn. (\ref{rho3local}) . Hence we have proved that 
$(I_A \otimes \$_B) (|\Phi^+\rangle \langle\Phi^+|)$ is a local state if  $(I_A \otimes \$_B)(\sigma_{AB})$ is a local state and
  $Tr_B(\sigma_{AB})= \frac{I_2}{2}$ .

\hfill $\square$

\end{document}